\documentclass[12pt,a4paper]{article}
\usepackage[utf8]{inputenc}
\usepackage{amsmath,amsfonts,amssymb,amsthm,textcomp}
\usepackage{amsthm}
\usepackage{hyperref}
\usepackage[font=footnotesize,labelsep=period]{caption}
\usepackage{float}
\usepackage{multirow}
\usepackage{multicol}
\usepackage{graphicx}
\usepackage[margin=2.9 cm]{geometry}
\usepackage{titlesec}
\titlelabel{\thetitle.\quad}

\begin{document}

\centerline{\Large \bf Integrability and rational soliton solutions for gauge}
\vskip 0.35cm
\centerline{\Large \bf invariant derivative nonlinear Schr\"odinger equations}
\vskip 0.7cm

\centerline{Paz Albares\footnote{Based on the contribution presented at the ``Third BYMAT Conference: Bringing Young Mathematicians Together", December 1--3, 2020, Valencia, Spain. To appear in the Proceedings of the Third BYMAT Conference.}}
\vskip 0.5cm
\centerline{Departamento de F\'isica Fundamental, Universidad de Salamanca, Spain}
\centerline{paz.albares@usal.es}

\vskip 0.7cm

\begin{abstract}
\noindent
The present work addresses the study and characterization of the integrability of three famous nonlinear Schr\"{o}dinger equations with derivative-type nonlinearities in $1+1$ dimensions. Lax pairs for these three equations are successfully obtained by means of a Miura transformation and the singular manifold method. After implementing the associated binary Darboux transformations, we are able  to construct rational soliton-like solutions for those systems.  
\end{abstract}


\vskip 0.5cm

\noindent
{\bf MSC class}: 35C08, 35Q55, 37J35
\vskip 0.3cm


\vskip 0.5cm

\section{Introduction}\label{sec:1}
The nonlinear Schr\"odinger (NLS) equation is one of the most famous integrable equations in soliton theory and mathematical physics \cite{ablowitz1991}. Among the several integrable generalizations of NLS, we are interested in the study of modified NLS systems with derivative-type nonlinearities in $1+1$ dimensions, which are known as derivative nonlinear Schr\"odinger (DNLS) equations. There exist three celebrated equations of this kind, {\it i. e.} the Kaup-Newell (KN) system \cite{kaup1978},
\begin{equation}
im_t-m_{xx}-i\left(\left|m\right|^2m\right)_x=0
\label{eq:KN}
\end{equation}
the Chen-Lee-Liu (CLL) equation \cite{chen1979},
\begin{equation}
im_t-m_{xx}-i\left|m\right|^2m_x=0
\label{eq:CLL}    
\end{equation}
and the Gerdjikov-Ivanov (GI) equation \cite{gerdjikov1983}
\begin{equation}
im_t-m_{xx}+im^2\overline{m}_x-\frac{1}{2}\left|m\right|^4m=0
\label{eq:GI}
\end{equation}
where $m$ is a complex valued function and $\overline{m}$ denotes the complex conjugate of $m$. 

It is already known that these three equations are equivalent via a $U(1)$-gauge transformation \cite{kundu1984}. If $m(x,t)$ is a solution of the KN system \eqref{eq:KN}, it is easy to find that the new field $M(x,t)$
\begin{equation}
M(x,t)=m(x,t)\,e^{\frac{i\gamma}{2}\,\theta(x,t)},\quad \text{with}\quad \theta_x=\left|m\right|^2,\quad \theta_t=i(m\overline{m}_x-\overline{m}m_x)+\frac{3}{2}\left|m\right|^4   
\label{eq:4}
\end{equation}
satisfies the CLL equation for $\gamma=1$, and the GI equation for $\gamma=2$. 

Gauge transformations constitute an useful tool to link integrable evolution equations in soliton theory, since they provide B\"acklund transformations between those equations as well as the relation of their associated linear problems \cite{wadati1983}. 
In this contribution we exploit this gauge invariance property to construct a Lax pair and rational soliton solutions for these three equations. For a detailed analysis and explicit calculations, we refer the reader to \cite{albares2020}. 

\section{Integrability and Lax pair} \label{sec:2}
The Painlev\'e test \cite{weiss1983} has been proved to be a powerful criterion for the identification of integrable partial differential equations (PDEs). A PDE is said to posses the Painlev\'e property, frequently considered as a proof of integrability, when its solutions are singled-valued about the movable singularity manifolds. This requires the generalized Laurent expansion for the field $m(x,t)=\sum_{j=0}^\infty a_j(x,t)\,\phi(x,t)^{j-\mu}$, where $\phi(x,t)$ is an arbitrary function called the singular manifold and the index $\mu\in\mathbb{N}$ is an integer. 

The Painlev\'e test is unable to check the integrability of any DNLS equation since the leading index is not integer, $\mu=\frac{1}{2}$. This fact allow us to introduce two new real fields $\alpha(x,t),\, \beta(x,t)$
\begin{equation}
m(x,t)=\sqrt{2\alpha_x}e^{\frac{i}{2}\beta(x,t)},\quad \text{with}\quad\alpha_x=\frac{1}{2}\left|m\right|^2,\quad \beta=(2\gamma-3)\alpha+\int{\frac{\alpha_t}{\alpha_x}dx}
\label{eq:5}    
\end{equation}
with $\gamma=0$ for the KN system, $\gamma=1$ for the CLL equation and $\gamma=2$ for the GI equation. This ansatz yields an identical differential equation for $\alpha$ in each case, expressed in the conservative form
\begin{equation}
\left[\alpha_x^2-\alpha_t\right]_t=\left[\alpha_{xxx}+\alpha_x^3-\frac{\alpha_t^2+\alpha_{xx}^2}{\alpha_x}\right]_x 
\label{eq:alpha}
\end{equation}
From expression \eqref{eq:4}, it can be easily seen that the probability density $\theta_x=\left|m\right|^2=\left|M\right|^2$ is invariant under a $U(1)$-gauge transformation, indeed it constitutes the first conservation law for these systems. Due to this symmetry, it is straightforward to see that once we obtain a soliton solution for a particular DNLS equation, it is immediate to derive soliton solutions for any DNLS equation linked by a $U(1)$-gauge transformation. 

Since $\alpha_x=\frac{\theta_x}{2}$, we may conclude that equation \eqref{eq:alpha} is the representative equation for the probability density of any DNLS equation. Equation \eqref{eq:alpha} passes the Painlev\'e test, but it possesses two branches of expansion. The best method to overcome this inconvenience requires the splitting of the field $\alpha$ as
\begin{equation}
\alpha=i(u-\overline u),\qquad\qquad \alpha_x^2-\alpha_t=u_{xx}+\overline u_{xx}
\label{eq:7}
\end{equation}
The combination of equations in \eqref{eq:7} yields two Miura transformations for $\{u,\,\overline{u}\}$ and the coupling condition
\begin{equation}
\begin{aligned}
u_{xx}=\frac{1}{2}&\left(\alpha_x^2-\alpha_t -i\alpha_{xx}\right),\qquad\overline u_{xx}=\frac{1}{2}\left(\alpha_x^2-\alpha_t +i\alpha_{xx}\right),\\[.3cm]
&iu_t+u_{xx}-i\overline u_t+\overline u_{xx}+(u_x-\overline u_x)^2=0
\end{aligned}
\label{eq:8}    
\end{equation}
which finally lead to the nonlocal Boussinesq equation \cite{lambert1994} for $u(x,t)$ of the form
\begin{equation}
\left[u_{tt}+u_{xxxx}+2u_{xx}^2-\frac{u_{xt}^2+u_{xxx}^2}{u_{xx}}\right]_x=0
\label{eq:u}
\end{equation}
where it may be easily checked that $\overline{u}(x,t)$ satisfies the same equation. Equation \eqref{eq:u} has the Painlev\'e property with an unique branch of expansion. Hence, this equation turns out to be integrable in the Painlev\'e sense and it is possible to derive an equivalent linear spectral problem associated to the nonlinear equation \eqref{eq:u}. This aim may be achieved by means of the so-called singular manifold method (SMM). 

The SMM \cite{weiss1983} focuses on solutions which emerge from the truncated Painlev\'e series, and act as auto-B\"acklund transformations, of the form $u ^{[1]}=u^{[0]}+\log(\phi)$. Thus, the singular manifold $\phi$ is no longer an arbitrary function, since it satisfies the singular manifold equations. The associated linear problem arises from the linearization of these equations, and it can be demonstrated that the Lax pair for $u$ reads \cite{albares2020}
\begin{equation}
    \begin{aligned}
    \psi_{xx}&=\left(\frac{u^{[0]}_{xxx}-iu^{[0]}_{xt}}{2u^{[0]}_{xx}}-i\lambda\right)\,\psi_x-u^{[0]}_{xx}\psi,\quad \psi_t=i\psi_{xx}-2\lambda\psi_x+i\left(2u^{[0]}_{xx}+\lambda^2\right)\psi\\
   \chi_{xx}&=\left(\frac{u^{[0]}_{xxx}+iu^{[0]}_{xt}}{2u^{[0]}_{xx}}+i\lambda\right)\,\chi_x-u^{[0]}_{xx}\chi,\quad \chi_t=-i\chi_{xx}-2\lambda\chi_x-i\left(2u^{[0]}_{xx}+\lambda^2\right)\chi\\
    \end{aligned}
    \label{eq:10}
\end{equation}
where $\{\chi,\psi\}$ are two complex conjugated eigenfunctions satisfying $\frac{\psi_x\chi_x}{\psi\chi}+u^{[0]}_{xx}=0$ and $\lambda$ is the spectral parameter. From \eqref{eq:10}, we may compute the Lax pair for the DNLS equations, obtaining 
\begin{equation}
\begin{aligned}
\chi_{xx}&=\left[i\lambda-\frac{i(\gamma-2)}{2}\left|m^{[0]}\right|^2+\frac{m^{[0]}_x}{m^{[0]}}\right]\chi_x+\frac{1}{2}\left[im^{[0]}{\overline{m}^{[0]}_x}-\frac{\gamma-1}{2}\left|m^{[0]}\right|^4\right]\chi\\
\chi_{t}&=i\chi_{xx}-\left[(\gamma-2)\left|m^{[0]}\right|^2+\frac{2im^{[0]}_x}{m^{[0]}}\right]\chi_x-i\lambda^2\chi
\label{eq:11}
\end{aligned}
\end{equation}
and its complex conjugate, for the corresponding value of $\gamma$ in each case. It is worthwhile to remark that the coupling condition for the Lax pair in $u$ gives rise to
$\frac{\psi_x\chi_x}{\psi\chi}-\frac{i}{2}\,m^{[0]}\overline{m}^{[0]}_x+\frac{\gamma-1}{4}\left|m^{[0]}\right|^4=0$, which allows us to determine an additional but completely equivalent Lax pair for those systems.

\section{Rational soliton solutions} \label{sec:3}
Once the Lax pair have been obtained for a given PDE by means of the SMM, binary Darboux transformations can be constructed in order to obtain iterated solutions for that PDE. We implement the Darboux transformation formalism over the spectral problem \eqref{eq:10} so as to provide a general iterative procedure to compute up to the $n$th iteration for $u$. By virtue of expressions \eqref{eq:4}, \eqref{eq:5} and \eqref{eq:7}, solutions for the DNLS equations can be forthrightly established. Thus, soliton solutions for DNLS equations may be derived by considering a suitable choice for the seed solution and the eigenfunctions in the Lax pair.

In the following lines we summarize the main results regarding this procedure, oriented to the obtention of rational soliton solutions. Further details and a general rigorous analysis may be found in \cite{albares2020}.

\begin{figure}[H]
\centering
\includegraphics[width=0.47\textwidth]{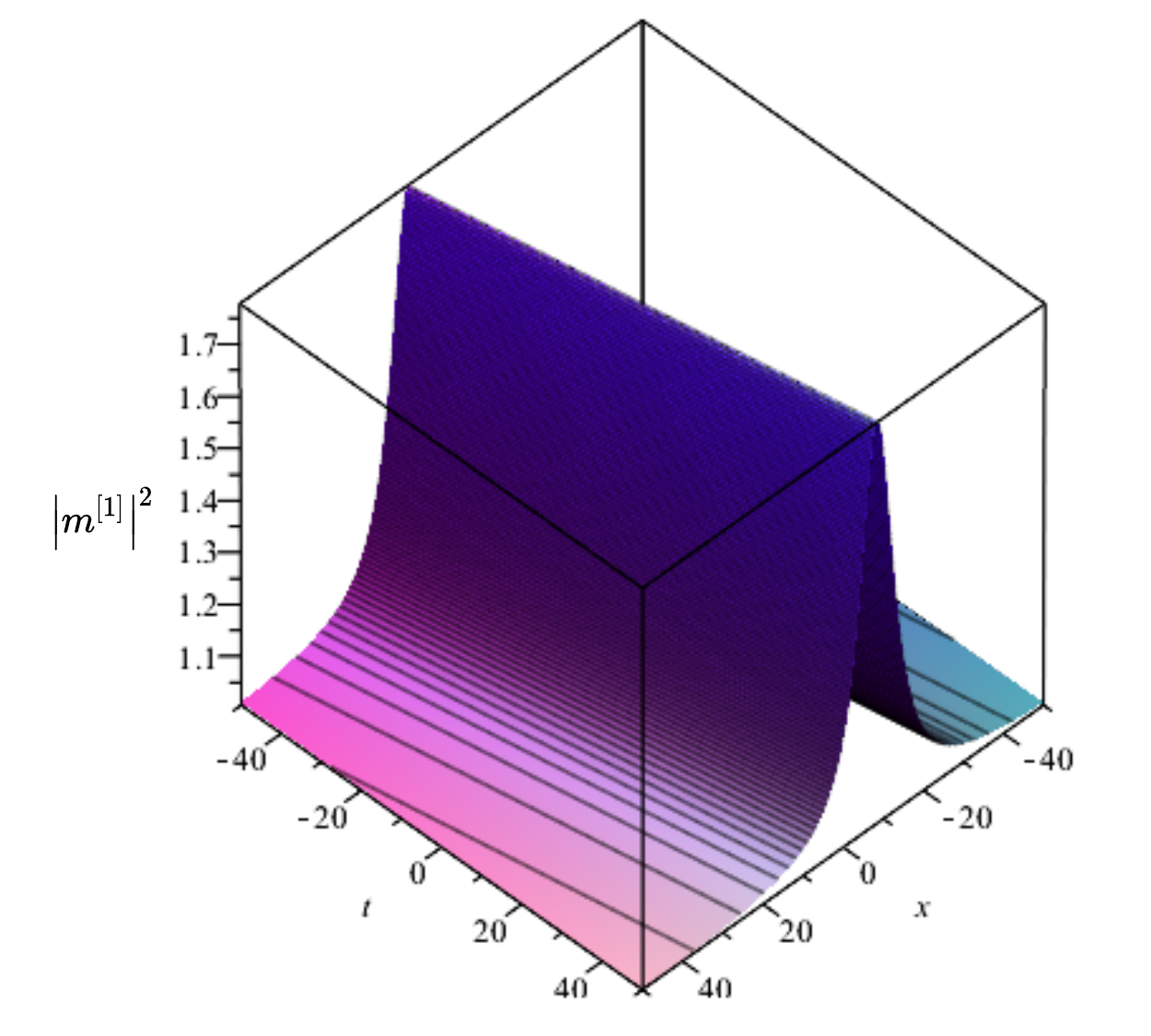}
\hspace{0.03\textwidth}
\includegraphics[width=0.47\textwidth]{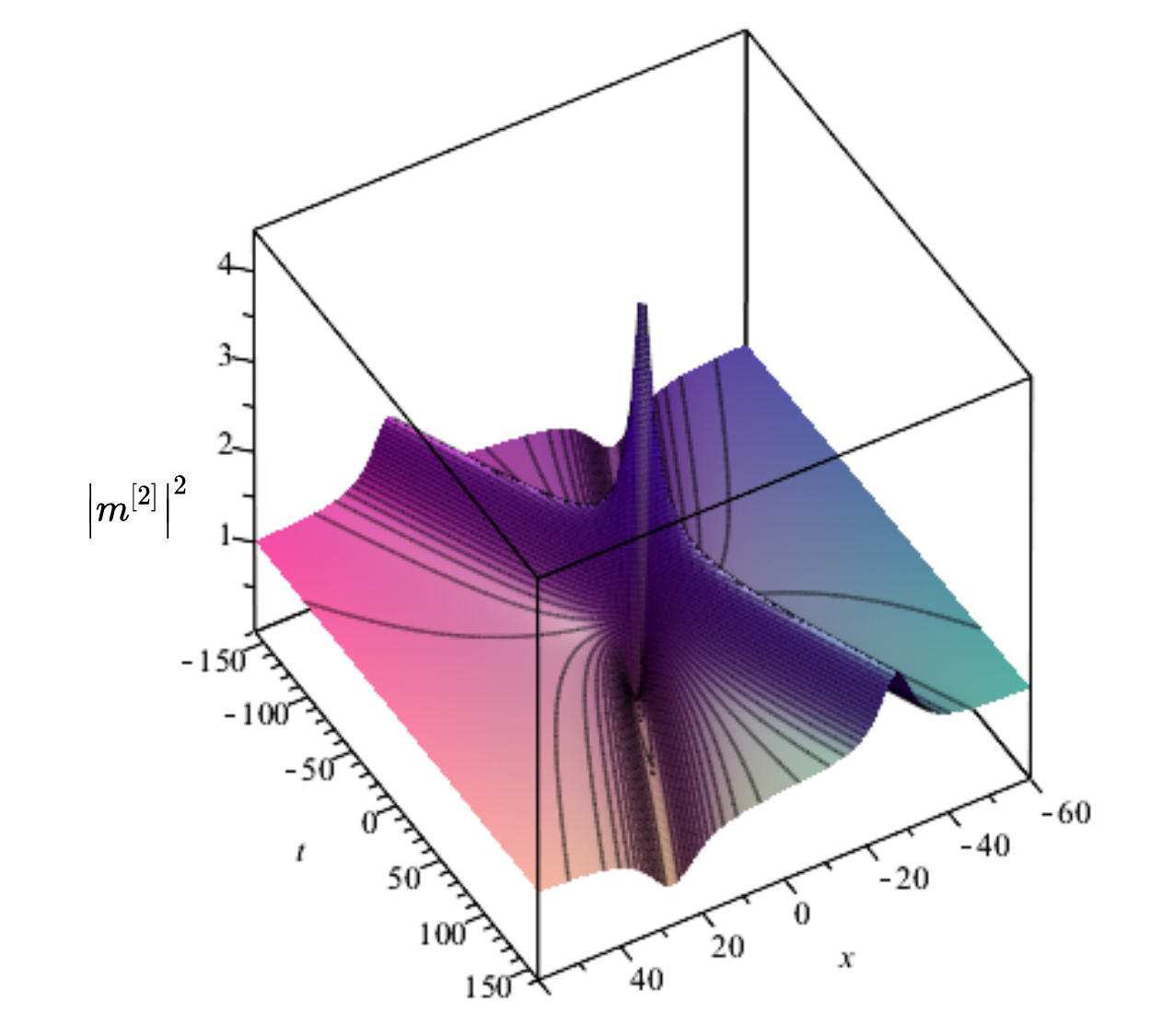}
\caption{Spatio-temporal plot of $\left|m^{[1]}\right|^2$ and $\left|m^{[2]}\right|^2$ for parameters $\sigma=-1,\,j_0=1,\,z_0=\frac{1}{6}$.}
\label{fig:1}
\end{figure}

We start from a polynomial seed solution $u^{[0]}$ for \eqref{eq:u} and binary exponential eigenfunctions for \eqref{eq:10},
\begin{equation}
\begin{aligned}
u^{[0]}&=-\frac{j_0^2}{4}\left[j_0^2z_0^2x\left(\frac{x}{2}+j_0^2(z_0^2+1)\,t\right)+i\left(x+j_0^2\left(z_0^2+\frac{1}{2}\right)t\right)\right],\\[.3cm]
\chi_{\sigma}&=e^{\,\,\frac{i}{2}j_0^2z_0\sigma\,\left[x+j_0^2\left(-\frac{\sigma}{2z_0}(z_0^4+7z_0^2+1)+3(z_0^2+1)\right)t\right]},\qquad \psi_{\sigma}=\overline{\chi}_{\sigma}
\label{eq:12}
\end{aligned}
\end{equation}
where $j_0$ and $z_0$ are arbitrary parameters, $\sigma=\pm 1$ and $\lambda_{\sigma}=\frac{j_0^2}{2}\left(2\sigma z_0-(z_0^2+1)\right)$. The first and second iterations $u^{[j]},\,j=1,2$ can be performed and the soliton solution profile may be computed as $\left|m^{[j]}\right|^2=2i(u^{[j]}_x-\overline{u}^{[j]}_x)$. The results are displayed in Figure \ref{fig:1}.

The first iteration ($j=1$) provides a rational soliton-like travelling wave along the $x-j_0^2\left(\sigma z_0-(z_0^2+1)\right)t$ direction and constant amplitude, of expression
\begin{equation}
\left|m^{[1]}_{\sigma}\right|^2=j_0^2-\frac{4}{j_0^2z_0(\sigma-z_0)\left[\left(x-j_0^2\left(\sigma z_0-(z_0^2+1)\right)t\right)^2+\frac{1}{j_0^4z_0^2(\sigma-z_0)^2}\right]},\quad \sigma=\pm 1
\label{eq:13}
\end{equation}
For the second iteration ($j=2$), we get the two-soliton solution
\begin{equation}
\left|m^{[2]}\right|^2=j_0^2
+\frac{8\left[\left(x+j_0^2(z_0^2+2)\,t\right)^2+j_0^4(z_0^2-1)\,t^2+\frac{1}{j_0^4(z_0^2-1)}\right]}{j_0^2(z_0^2-1)\left[\left(\left(x+j_0^2(z_0^2+1)\,t\right)^2-j_0^4z_0^2t^2-\frac{1}{j_0^4(z_0^2-1)}\right)^2+\frac{4\left(x+j_0^2(z_0^2+2)\,t\right)^2}{j_0^4(z_0^2-1)^2}\right]}
\label{eq:14}
\end{equation}
leading to a two asymptotically travelling rational solitons of the form \eqref{eq:13} (for $\sigma=1$ and $\sigma=-1$, respectively) interacting at the origin.


\begin{thebibliography}{99}
\bibitem{ablowitz1991} M. A. Ablowitz and P. A. Clarkson, \textit{Solitons, Nonlinear Evolution Equations and Inverse Scattering}, Cambridge University Press (1991)
\bibitem{kaup1978} D. Kaup and A. C. Newell, An exact solution for a derivative nonlinear Schr\"odinger equation, {\it J. Math. Phys.} \textbf{19}, 798-–801 (1978)
\bibitem{chen1979} H. H. Chen, Y. C. Lee and C. S. Liu, Integrability of Nonlinear Hamiltonian Systems by Inverse Scattering Method, {\it Phys. Scr.} \textbf{20}, 490--492 (1979)
\bibitem{gerdjikov1983} V. Gerdjikov and I. Ivanov, A quadratic pencil of general type and nonlinear evolution equations. II. Hierarchies of Hamiltonian structures, {\it Bulg. J. Phys.} \textbf{10}, 130--143 (1983)
\bibitem{kundu1984} A. Kundu, Landau-Lifshitz and higher-order nonlinear systems gauge generated from nonlinear Schr\"odinger-type equations, {\it J. Math. Phys.} \textbf{25}, 3433--3438 (1984)
\bibitem{wadati1983} M. Wadati and K. Sogo, Gauge Transformations in Soliton Theory, {\it J. Phys. Soc. Jpn.} \textbf{52}, 394--398 (1983)
\bibitem{albares2020} P. Albares, P. G. Est\'evez and J. D. Lejarreta, Derivative non-linear Schr\"odinger equation: Singular manifold method and Lie symmetries, {\it Appl. Math. Comput.} \textbf{400}, 126089 (2021)
\bibitem{weiss1983} J. Weiss, The Painlev\'e property for partial differential equations. {II}: B\"acklund transformation, Lax pairs, and the Schwarzian derivative,  {\it J. Math. Phys.} \textbf{24}, 1405–-1413 (1983)
\bibitem{lambert1994} E. Lambert, I. Loris, J. Springael and R. Willer, On a direct bilinearization method: Kaup{\textquotesingle}s higher-order water wave equation as a modified nonlocal Boussinesq equation, {\it J. Phys. A: Math. Gen.} \textbf{27}, 5325--5334 (1994)
\end{thebibliography}
\end{document}